\documentclass[%
 aip,
 amsmath,amssymb,
 reprint,%
]{revtex4-1}

\usepackage{graphicx}
\usepackage{dcolumn}
\usepackage{bm}

\usepackage[utf8]{inputenc}
\usepackage[T1]{fontenc}
\usepackage{mathptmx}

\begin{document}

\preprint{AIP/123-QED}

\title[Recurrence Networks from Light Curves]{Classification of Close Binary Stars Using Recurrence Networks}

\author{Sandip V. George}
\email{sandip.varkey@students.iiserpune.ac.in}
\affiliation{ 
Indian Institute of Science Education and Research (IISER) Pune, Pune - 411008, India
}%
 \altaffiliation[Also at ]{Interdisciplinary Center Psychopathology and Emotion Regulation, University of Groningen, University Medical Centre of Groningen, Groningen, The Netherlands}
\author{R. Misra}%
 \email{rmisra@iucaa.in}
\affiliation{ 
Inter University Centre for Astronomy and Astrophysics (IUCAA), Pune - 411007, India
}%

\author{G. Ambika}
 \email{g.ambika@iisertirupati.ac.in}
\affiliation{Indian Institute of Science Education and Research (IISER) Tirupati, Tirupati-517507, India}%

\date{\today}

\begin{abstract}
Close binary stars are binary stars where the component stars are close enough such that they can exchange mass and/or energy. They are subdivided into semi-detached, overcontact or ellipsoidal binary stars. A challenging problem in the context of close binary stars, is their classification into these subclasses, based solely on their light curves. Conventionally, this is done by observing subtle features in the light curves like the depths of adjacent minima, which is tedious when dealing with large datasets. In this work we suggest the use of machine learning algorithms applied to measures of recurrence networks and nonlinear time series analysis to differentiate between classes of close binary stars. We show that overcontact binary stars occupy a region different from semi-detached and ellipsoidal binary stars in a plane of characteristic path length(CPL) and average clustering coefficient(CC), computed from their recurrence networks. We use standard clustering algorithms and report that the clusters formed corresponds to the standard classes with a high degree of accuracy.
\end{abstract}

\maketitle

\begin{quotation}
Our study aims to classify close binary stars into semi-detached, overcontact and ellipsoidal binaries based on the properties of the recurrence networks constructed from their light curves. We show how this can be automated using machine learning algorithms for faster and efficient classification. Recurrence networks have been an important addition to conventional nonlinear time series analysis tools to study the dynamical behavior of real world systems, with a major advantage in terms of the number of data points required for reliable analyses. The methods of machine learning are especially suited to deal with large numbers of astrophysical objects. In this context, study of nonlinear dynamics of binary star systems has been mostly restricted to compact objects like neutron stars and black holes, where matter accretion leads to interesting dynamical phenomena\cite{misra2004chaotic,karak2009search}. On the other hand, nonlinear time series analysis of noncompact binary stars has not been studied in detail. In close binary stars, exchange of matter and energy leads to irregularities in their light curves and these irregularities reflect the variations in their underlying dynamics. We start with methods of nonlinear time series analysis to recreate the dynamics and capture the subtle variations in their dynamics by constructing recurrence networks from them. Then clustering algorithms like k-means and support vector machines are used to see the pattern of clusters in the plane of characteristic path length(CPL) and average clustering coefficient(CC), which can correctly identify the three categories of the stars. Our method is computationally much less expensive than conventional methods and can be effective with much smaller data sets and hence an efficient way to deal with a large number of data sets.
\end{quotation}

\section{\label{sec:intro}Introduction}
\begin{figure}
\begin{center}
\includegraphics[width=.9\columnwidth]{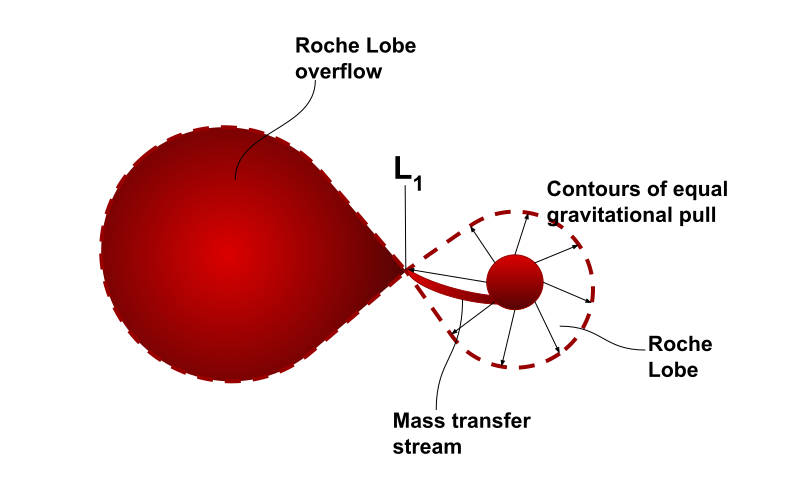}
\end{center}
\caption{\label{fig:sch} Schematic of a semi-detached binary star showing the Roche lobes, the inner Lagrangian point and the mass transfer stream between the stars.}
\end{figure}

Binary stars are systems of two stars which are gravitationally bound to each other. When the stars are aligned such that their movement results in eclipses from the earth's line of sight, they are termed as eclipsing binaries. Morphologically they may be classified into detached, semi-detached and overcontact binary stars\cite{kallrath2009eclipsing}. They are formally defined using the concept of Roche lobes, which constitutes the region around a star where its matter is gravitationally tied to it. When both of the stars in the binary do not fill their Roche lobes, the binary is termed as a detached binary. When only one of the stars fill its Roche lobe, it is called a semi-detached binary and when both components fill their Roche lobes, it is called an overcontact binary star\cite{shu1982physical}. A schematic of a semi-detached binary star is shown in Figure \ref{fig:sch}. The term close binary stars is used to collectively refer to the latter two configurations where mass and/or energy transfer is possible. This happens through the inner Lagrangian point, $L_1$ as shown in Figure \ref{fig:sch}. If the stars are inclined so that no eclipse is visible from our line of sight, we categorize the stars as ellipsoidal binary stars. In this case, the mutual gravitation between the stars distorts their shapes, which is the primary cause of their light variation\cite{morris1985ellipsoidal}. Hence, the underlying dynamical processes that govern the variations in the light curves of the three classes of close binaries are different. In this work, we study how the nonlinear quantifiers computed from reconstructed dynamics from the light curves of all the close binary stars in the Kepler field of view, can be used to classify them into different categories\cite{prvsa2011kepler,slawson2011kepler}.

The framework of recurrence networks have been put to extensive use to study real world complex systems from time series or observational data. It is proved to be useful in differentiating between dynamical states, studying dynamical transitions etc\cite{marwan2009complex,gao2013recurrence,godavarthi2017recurrence}. The plane of the characteristic path length and the average clustering coefficient has been shown to be useful in differentiating between different dynamical states that can be exhibited by a system\cite{jacob2016characterization}. 
Light variability in many astrophysical sources have been shown to be due to nonlinear dynamics and chaos\cite{buchler1995chaotic,mindlin1998low,plachy2018chaotic}. Hence we are motivated to use the measures of recurrence network to discern the differences in underlying dynamics of binary stars and use that information in classifying them into proper groups.  Our classification is then cross checked with the existing astrophysical classification to validate our results.
The relevance of the study lies in the fact that differentiating between the different categories of close binaries is really challenging. This is especially true when distinguishing between ellipsoidal and overcontact binary stars, which are often found at similar inclinations. Traditionally, morphological classification is conducted for binary stars by careful examination of the features of the light curves. This becomes impractical and computationally expensive with increasing sizes of datasets. Since the underlying phenomena responsible for the light variations in the different categories of close binary stars are different, we expect different values for the nonlinear quantifiers from the corresponding light curves.

We reconstruct the phase space dynamics from the light curves of the close binaries and then recurrence networks are constructed from the trajectory or attractor in the phase space. The properties of the network thus constructed reflects the pattern of recurrence of phase points on the attractor and hence the geometrical properties of the underlying  dynamics. We focus on two main measures computed from the recurrence networks, namely the characteristic path length and average clustering coefficient. We use clustering algorithms on this CPL-CC plane and explore the clusters that are formed in it. We compare these clusters with the existing classification schemes to check how well these correlate with each other.  

Our study thus presents an alternate classification scheme using recurrence networks from observational data which is further made effective and automated using clustering and machine learning algorithms. We show that the classification scheme developed agrees with the classification done previously, with high accuracy.

\section{\label{sec:rnfromlc}Recurrence Networks Constructed from Light Curves}

\begin{figure}
\begin{center}
\includegraphics[width=.9\columnwidth]{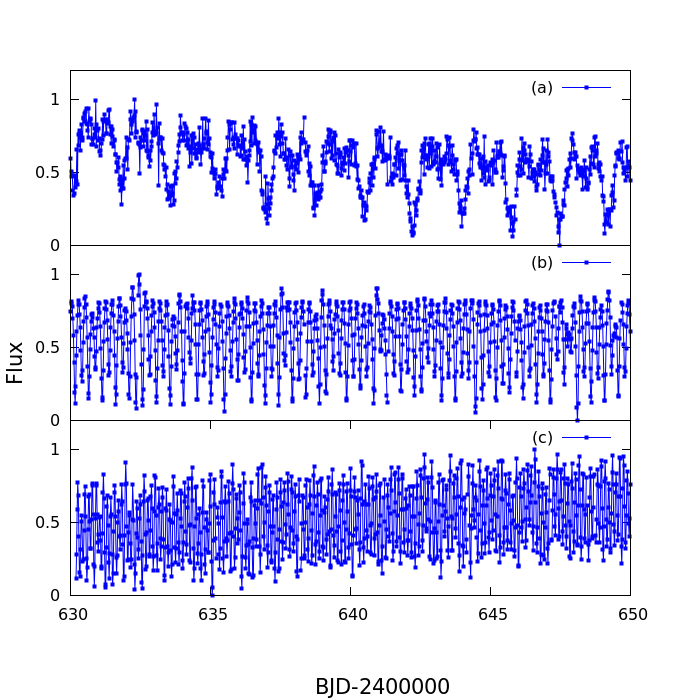}
\end{center}
\caption{\label{fig:closelc} Typical light curves of (a)Semi-detached star KIC 8509469 (b)Overcontact star KIC 10796477 and (c) Ellipsoidal star KIC 5564600.}
\end{figure}

Our dataset consists of close binary stars in the Kepler field of view, classified as such in the second revision of the Kepler eclipsing binary catalog\cite{slawson2011kepler}. The catalogue divided the eclipsing binaries into semi-detached, overcontact and ellipsoidal morphological classes, containing $152$, $463$ and $137$ samples respectively. Typical light curves from these three categories of close binary stars are shown in Figure \ref{fig:closelc}. We use only the first $3000$ points of a light curve, which is initially cleaned for outlier points using a standard deviation estimated from the median (the median absolute deviation)\cite{chen1993joint,leys2013detecting}. These data sets are then subjected to preliminary analysis that involves conversion into uniform deviates. This transformation rescales all of them uniformly  into the range [0:1], with uniform distributions. Then the data is embedded in a phase space of chosen dimension by using the method of delay embedding\cite{takens1981detecting}. The value at which the autocorrelation falls to $\frac{1}{e}$ is chosen as the required delay for each dataset \cite{rosenstein1994reconstruction,harikrishnan2006non}. If $I_u(t)$ is the value of the intensity received from the star at time $t$, after the light curve is converted into a uniform deviate and $\tau$ is the delay time, a reconstructed vector in a phase space of dimension $M$ is given by,
\begin{equation}
    \Vec{v}_i=[I_u(t_i), I_u(t_i+\tau), ..., I_u(t_i+(M-1)\tau)]
\end{equation}
From this reconstructed space, the recurrence network is generated by choosing a threshold $\epsilon$ such that attractor points that recur within, or are closer than, $\epsilon$ are connected. For this we generate the binary recurrence matrix $R_{i,j}(\epsilon)$ as
\begin{equation}
R_{i,j}(\epsilon)=\Theta(\epsilon-||v_i-v_j||)
\end{equation}
Here, $\Theta$ is the Heaviside step function and $||.||$ is chosen as the Euclidean norm\cite{donner2010recurrence}. Each point on the reconstructed attractor serves as a node of the network with a link present when $R_{i,j}$ is $1$. The pattern of connections in the network thus generated is captured in the matrix of connections called an adjacency matrix $A_{i,j}$, computed from the recurrence matrix by setting the diagonal elements to zero\cite{donner2010recurrence}. 
\begin{equation}
A_{i,j}(\epsilon)=R_{i,j}(\epsilon)-\delta_{i,j}
\end{equation} 
where $\delta_{i,j}$ is the Kronecker delta function. 

\begin{figure}
\begin{center}
\includegraphics[width=.7\columnwidth]{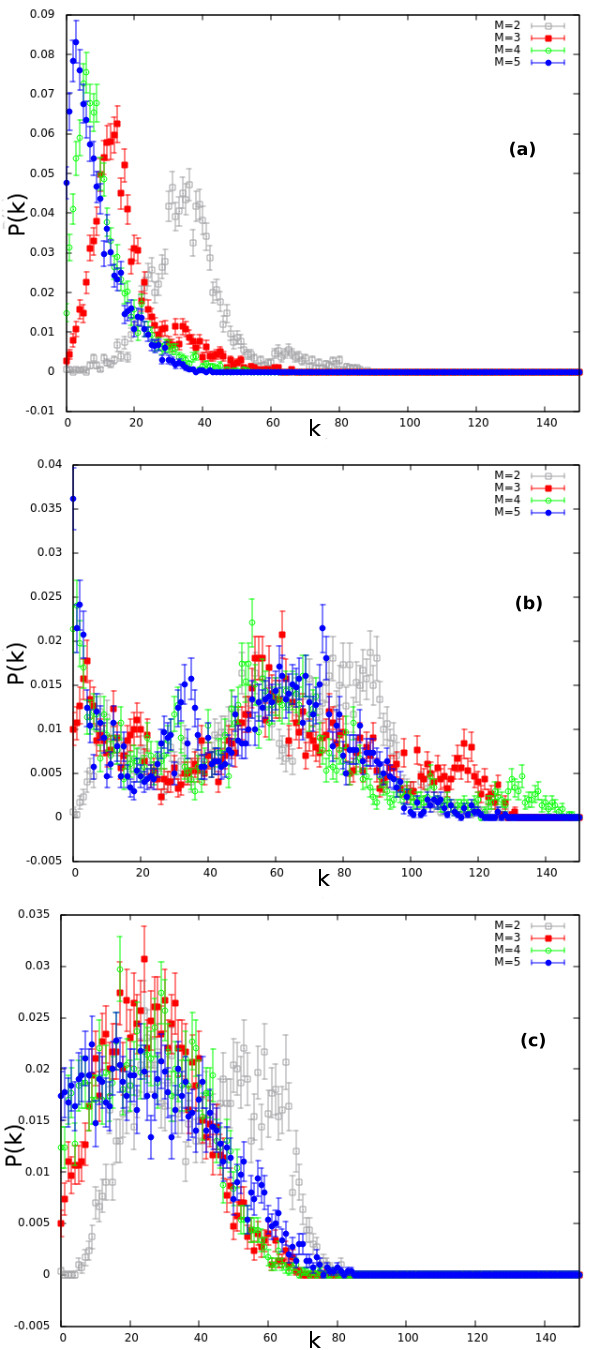}
\end{center}
\caption{\label{fig:degdist} Normalized degree distributions for differing embedding dimensions, $M$, for recurrence networks constructed from the light curves of (a)Semi-detached star KIC 8509469 (b)Overcontact star KIC 10796477 and (c) Ellipsoidal star KIC 5564600. All three show saturation of the distribution at $M=4$ or earlier. The error on a degree value, $k$, is given by $\frac{\sqrt{n(k)}}{N}$, where $n(k)$ is the number of nodes with degree $k$\cite{jacob2016uniform}. }
\end{figure}

$A_{i,j}$ is used to compute the characteristic measures of the network\cite{boccaletti2006complex}. 
An important first step in this procedure is the choice of a proper embedding dimension to reconstruct the attractor in phase space. For our analysis we use the embedding dimension $4$ for all the stars, since  in almost all cases considered, the degree distribution of the recurrence network saturates at $M=4$, as shown in Figure \ref{fig:degdist} \cite{jacob2016uniform}.

\begin{figure}
\begin{center}
\includegraphics[width=.7\columnwidth]{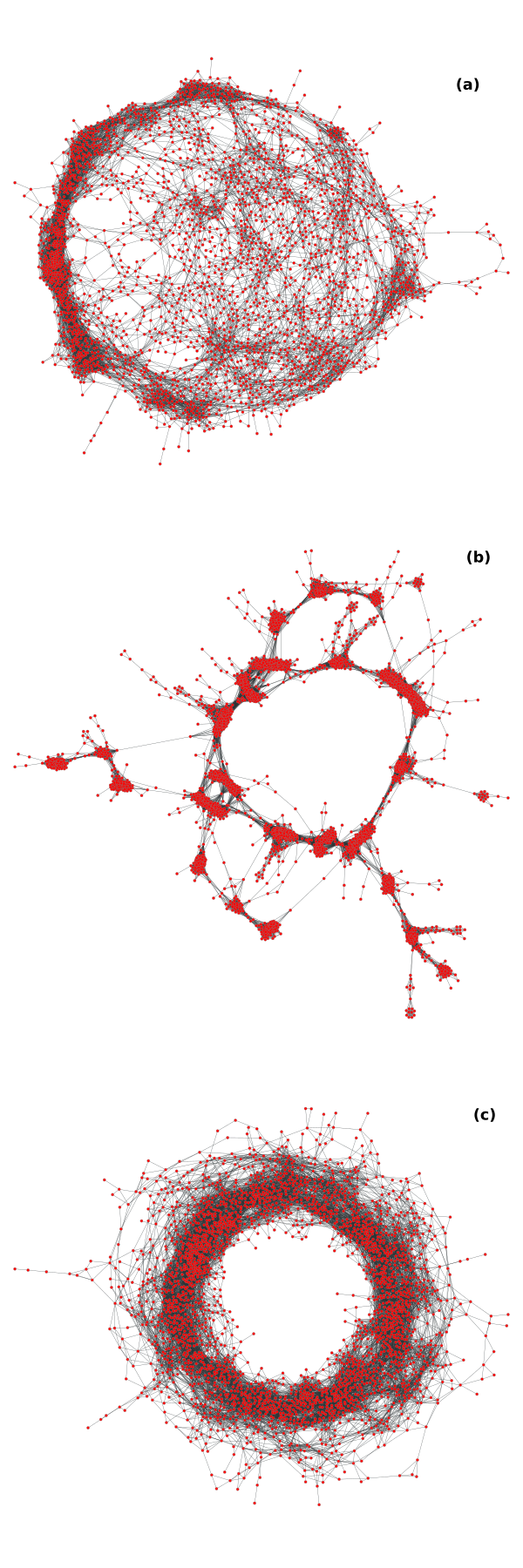}
\end{center}
\caption{\label{fig:recnet} Recurrence networks constructed from the light curves of (a)Semi-detached star KIC 8509469 (b)Overcontact star KIC 10796477 and (c) Ellipsoidal star KIC 5564600.}
\end{figure}

The next important choice is that of the recurrence threshold $\epsilon$. It is chosen such that at least $95\%$ of the nodes form part of the largest component in the network. For $M=4$, this threshold is set to be $\epsilon=0.14$, as suggested in Jacob et. al.\cite{jacob2016uniform}. The recurrence networks created using sample light curves from the three classes of close binary stars are shown in Figure \ref{fig:recnet}. 

Then we proceed to check whether the recurrence networks thus generated from the embedded attractor, show distinct features for different classes of close binaries. This requires quantification of the recurrence networks using two main characteristic measures, namely the characteristic path length and the clustering coefficient. The characteristic path length (CPL) is measured as the average of shortest paths between every possible node pair $(i,j)$\cite{boccaletti2006complex}. 
\begin{equation}
    CPL = \frac{1}{N \cdot (N - 1)} \cdot \sum_{i \ne j} d_{i,j}
\end{equation}
Here $i$ and $j$ are nodes in the network and $d_{i,j}$ is the shortest path between them. 
The local clustering for each node measures how close a node is to being a complete graph (or clique). If $N$ is total number of nodes in the network, the local clustering for a node $i$ having degree $k_i$ is given by\cite{jacob2016uniform,newman2018networks}
\begin{equation}
    C_i=\frac{\sum_{j,k}A_{ji}A_{jk}A_{ki}}{k_i(k_i-1)}
    \label{eq:locclus}
\end{equation}
The average clustering for the entire network is then given by\cite{newman2018networks}
\begin{equation}
    CC_{avg}=\frac{1}{n}\sum_{i=1}^{n}C_i
\end{equation}

\section{Classification and Clusters based on Measures of Recurrence Networks}
\subsection{\label{sec:cpl-cc based}CPL and CC values of close binary stars}
\begin{figure}
\begin{center}
\includegraphics[width=\columnwidth]{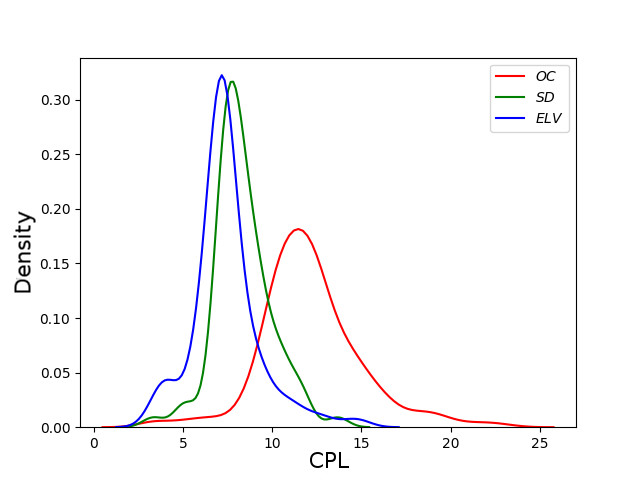}
\end{center}
\caption{\label{fig:hist_cpl_closebin}Distributions of Characteristic Path Lengths for the three kinds of close binary stars. The distributions for semi-detached and ellipsoidal binary stars are distinctly different from the distribution for the overcontact stars.}
\end{figure}

We try a classification scheme for close binary stars  based on the measures computed from their recurrence networks. We observe that the semi-detached and ellipsoidal binary stars have very different values for characteristic path length from overcontact binary stars. This is shown in Figure \ref{fig:hist_cpl_closebin} which shows the kernel density plot for the three categories of close binaries.

This would mean that using CPL alone, we should be able to identify the overcontact binary stars from semi-detached and ellipsoidal binary stars. From the kernel density plot shown in Figure \ref{fig:hist_cpl_closebin}, we define the intervals of CPL for the different categories. Thus at $CPL=9.8$, the distributions are found to merge for overcontact and demi-detached binary stars, whereas they merge at $CPL=9.2$ for overcontact and ellipsoidal stars. Using these intervals, we correctly classify $402$ of $463$ overcontact stars and $129$ of $152$ semi-detached stars. For the case of ellipsoidal binaries, $428$ of $463$ overcontact stars and $120$ of $137$ ellipsoidal stars are correctly classified. We define the accuracy of classification or prediction as the correctly predicted fraction of the total set. Thus for we can distinguish between semi-detached and overcontact binaries with an accuracy of $0.86$ while for classifying ellipsoidal stars and overcontact stars, an accuracy of $0.91$ is achieved.

\begin{figure}
\begin{center}
\includegraphics[width=\columnwidth]{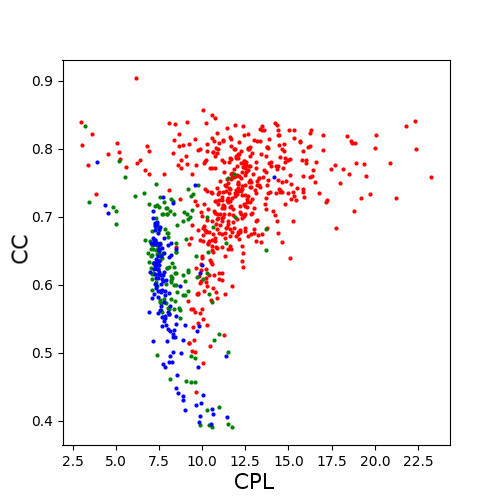}
\end{center}
\caption{\label{fig:cplccplane_closebin} Locations of close binary stars on the CPL-CC plane. Each point represents a particular close binary. Green corresponds to semi-detached, red corresponds to overcontact and blue to ellipsoidal binary stars. We notice that overcontact stars lie in a distinctly different part of the plane separated from the semi-detached and ellipsoidal stars.}
\end{figure}

We further look at both CPL and CC values of recurrence networks together.  In this context we note that the CPL-CC plane is known to have very different values for different dynamical states\cite{jacob2016characterization,jacob2017cross}. We try to locate all the close binary stars on the CPL-CC plane and it is shown in Figure \ref{fig:cplccplane_closebin}. Quite clearly, the semi-detached and ellipsoidal binary stars occupy a region of the plane very different from overcontact binary stars. In the following sections, we try to extract these clusters using one unsupervised learning algorithm, namely $k-means$ clustering and one supervised learning algorithm, namely support vector clustering.

\subsection{k-means clustering}
\begin{figure}
\begin{center}
\includegraphics[width=\columnwidth]{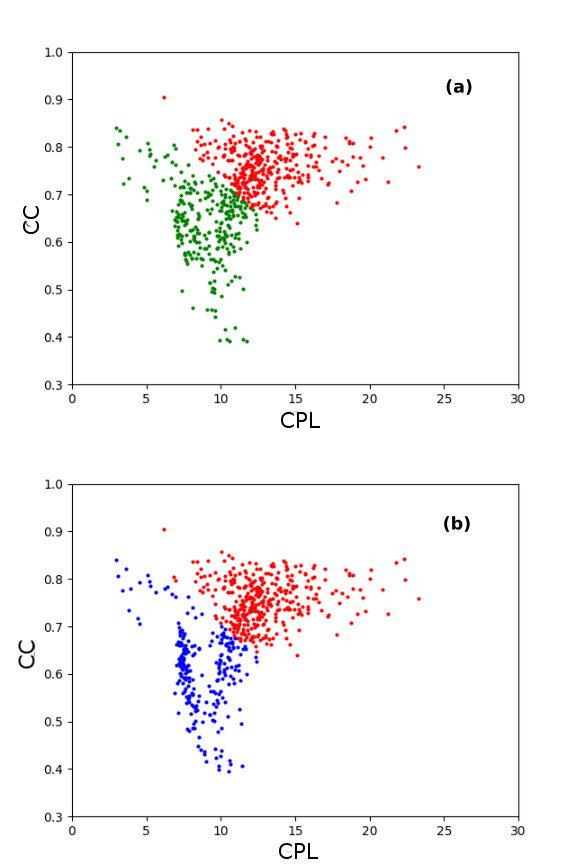}
\end{center}
\caption{\label{fig:kmeans_closebin} Clusters predicted by the $k-means$ clustering algorithm in the CPL-CC plane. Predicted clusters for (a)Overcontact and semi-detached stars and (b)Overcontact and ellipsoidal stars. The color code is green: semi-detached, red: overcontact and blue: ellipsoidal binary stars.}
\end{figure}

In this section, we show the use of a simple $k-means$ clustering algorithm to identify clusters in the $CPL-CC$ plane. The $k-means$ algorithm is an iterative algorithm that initially selects a set of $k$ random observations as initial $k-means$. Each observation is linked to the closest mean. The centroids of the clusters are calculated as the new $k-means$ and the process is repeated \cite{mackay2003information}. We initially attempt to distinguish between overcontact and semi-detached binary stars. We find the algorithm divides them into two regions in the CPL-CC plane (Figure \ref{fig:kmeans_closebin}a), correctly classifying $330$ of $463$ overcontact binaries and $146$ of $152$ semi-detached binaries. This leads to an accuracy of $0.77$. On considering ellipsoidal binaries, the clusters generated by the algorithm contains $357$ overcontact binaries and $135$ (of $137$) ellipsoidal binaries respectively (accuracy = $0.82$).
\subsection{Support vector machines}
We now report the use of support vector machines that can partition a multi dimensional distribution into clusters on the basis of a discriminating hyperplane. If we need to divide an $n-$ dimensional space into two clusters, the support vector machine constructs an optimal $n-1$ dimensional surface. This surface is decided so that the distance of the points closest to it, is maximized\cite{suykens1999least}. 

\begin{figure}
\begin{center}
\includegraphics[width=\columnwidth]{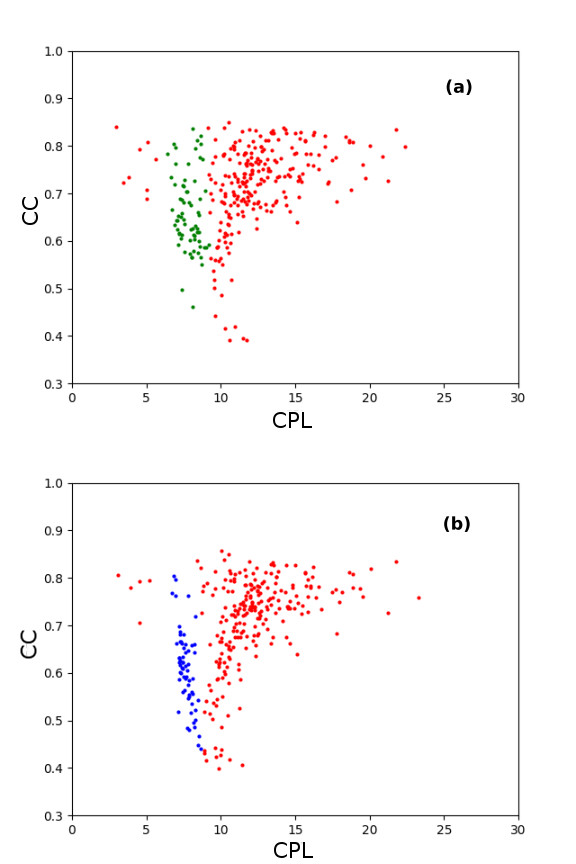}
\end{center}
\caption{\label{fig:svm_closebin} Clusters predicted by the support vector clustering algorithm in the CPL-CC plane. Predicted clusters for (a)Overcontact and semi-detached stars and (b)Overcontact and ellipsoidal stars. The color code is same as in Figure \ref{fig:kmeans_closebin}.}
\end{figure}

In our case, we need to partition the $2-d$ $CPL-CC$ plane into two clusters using a support vector machine. We keep aside $50\%$ of the points as the training set and the remaining $50\%$ of data as the testing set. As before, we initially differentiate between semi-detached and overcontact binaries. The algorithm gives an accuracy of $0.89\pm 0.01$ when applied to the testing set. In the case of ellipsoidal versus overcontact case, we get an accuracy of $0.91\pm 0.01$.  The error bars are standard deviations for ten iterations. The predicted clusters for a particular iteration is shown in Figure \ref{fig:svm_closebin}.

\begin{table}
\caption{Table giving the values of accuracy for different algorithms that differentiate overcontact stars from semi-detached and ellipsoidal stars.}
\begin{tabular*}{\columnwidth}{@{\extracolsep{\fill}}c c c} 
\hline\hline 
$Algorithm$  & $Accuracy(ELV)$ & $Accuracy(SD$)\\ [0.5ex] 
\hline 
$CPL$&$0.91$ & $0.86$\\
$k-means$ & $0.82$ & $0.77$\\ 
$SVC$  & $0.94\pm0.01$ & $0.89\pm0.01$\\ 
\hline 
\end{tabular*} 
\end{table}
\section{Summary and Discussion}

We report the results of our study on 752 close binary stars in the Kepler field of view using the framework of recurrence networks generated from their observational data. We invoke machine learning algorithms to distinguish between different types of close binary stars using the measures derived from their networks. Our study is based on the fact that the nonlinear dynamical properties of the light curves reflect the differences in the mechanisms for the variability among the three classes identified as semi-detached, overcontact and ellipsoidal binaries. We find that the range of values of characteristic path lengths(CPL) for the overcontact stars is different from ellipsoidal and semi-detached stars. Further the overcontact stars fall into a different region of the $CPL-CC$ plane, away from ellipsoidal and semi-detached binary stars. This indicates the possibility of organising them into clusters in the CPL-CC plane. We identify these clusters using two machine learning algorithms, one unsupervised and one supervised, the $k-means$ clustering algorithm and support vector machine. The clusters generated using these algorithms have a large intersection with the earlier astrophysical classifications. The accuracy of classification is remarkable but can be improved by further studies on proper choice of training sets. 

Our study illustrates, that the nonlinear dynamical properties of close binaries may be exploited to classify them into semi-detached, overcontact and ellipsoidal binary stars. This is because the variations in the underlying dynamics are reflected in the pattern of recurrences of the reconstructed attractors. We could quantify them using the measures of the corresponding recurrence networks. However, our results indicate that measures of the recurrence networks for ellipsoidal and semi-detached binary stars have similar range of values making it difficult to distinguish between the two categories. We think this must be because even though their inclinations are different, their underlying dynamics may have similar features. Hence our analysis which is based on the 
nature of nonlinear dynamical properties, may not distinguish them . However, using simple astrophysical parameters like the angle of inclination, their differentiation is easily possible\cite{prvsa2011kepler}.

With the challenges posed due to the need to classify large datasets like those from Kepler, methods such as local linear embedding have been suggested as alternatives\cite{matijevivc2012kepler}. The number of data points and computational power required for these techniques is larger than that for recurrence networks which give robust results with much smaller datasets. For instance all the computation done in this study used just 3000 data points per light curve. The good agreement with conventional classification suggests that this method is an efficient alternative for the classification using neural networks or local linear embedding. Further we have shown how we can refine the classification scheme by using machine learning algorithms which can be automated for use with very large number of data sets in a very efficient manner.

\end{document}